\documentclass[aps,prd,a4paper,onecolumn,amsmath,showpacs,superscriptaddress,nofootinbib,preprintnumbers,notitlepage]{revtex4-1}
 
\usepackage{verbatim}
\usepackage[T1]{fontenc}
\usepackage[utf8]{inputenc}
\usepackage[american]{babel}
\usepackage{epsfig}
\usepackage{graphicx,subcaption,caption}
\usepackage{pgfplots}
\usepackage{booktabs}
\usepackage{multirow}
\usepackage{dcolumn}
\usepackage{amsmath}
\usepackage{mathtools}
\usepackage{amsfonts}
\usepackage{amssymb}
\usepackage{epstopdf}
\usepackage{bm}
\usepackage{siunitx}
\usepackage{braket}
\usepackage{enumitem}
\usepackage{amsmath}

\usepackage{soul}
\usepackage{ulem}
\usepackage{color}
\usepackage{transparent}
\usepackage{pifont}
\usepackage[font={small}]{caption}

\definecolor{navyblue}{rgb}{0.0, 0.0, 0.5}
\definecolor{royalblue}{rgb}{0.25, 0.41, 0.88}
\definecolor{cadmiumgreen}{rgb}{0.0, 0.42, 0.24}
\definecolor{blue-violet}{rgb}{0.54, 0.17, 0.89}
\definecolor{darkviolet}{rgb}{0.58, 0.0, 0.83}
\definecolor{orange(colorwheel)}{rgb}{1.0, 0.5, 0.0}

\usepackage{hyperref}
\hypersetup{
    colorlinks=true, 
    linkcolor=royalblue, 
    citecolor=magenta}

\newcommand\be{\begin{equation}}
\newcommand\ee{\end{equation}}

\newcommand\bea{\begin{eqnarray}}
\newcommand\eea{\end{eqnarray}}

\usepackage{booktabs}
\usepackage{multirow}
\usepackage{dcolumn}
\usepackage{colortbl}

\definecolor{magenta(process)}{rgb}{1.0, 0.0, 0.56}

\definecolor{darkspringgreen}{rgb}{0.09, 0.45, 0.27}

\definecolor{royalblue(web)}{rgb}{0.25, 0.41, 0.88}

\begin{document}

\title{Constant-roll inflation with a complex scalar field}

\author{Ram\'{o}n Herrera}
\email{ramon.herrera@pucv.cl}
\affiliation{Instituto de F\'{\i}sica, Pontificia Universidad Cat\'{o}lica de Valpara\'{\i}so, Avenida Brasil 2950, Casilla 4059, Valpara\'{\i}so, Chile.
}

\author{Mehdi Shokri}
\email{mehdishokriphysics@gmail.com}
\affiliation{School of Physics, Damghan University, P. O. Box 3671641167, Damghan, Iran}
\affiliation{Department of Physics, University of Tehran, North Karegar Ave., Tehran 14395-547, Iran}
\affiliation{Canadian Quantum Research Center, 204-3002 32 Ave, Vernon, BC V1T 2L7, Canada}

\author{Jafar Sadeghi}
\email{pouriya@ipm.ir}
\affiliation{Department of Physics, University of Mazandaran, P. O. Box 47416-95447, Babolsar, Iran}

\preprint{}
\begin{abstract}
We consider inflation with a constant rate of rolling in which a complex scalar field plays the role of inflaton during the inflationary epoch. We implement the inflationary analysis for an accredited angular speed $\dot{\theta}$ which satisfies our dynamical equations. Scalar and tensorial perturbations generated in the framework of constant roll inflation with a complex field are studied. In this respect, we find  analytically solutions to the gauge invariant fluctuations, with which an expression for the scalar power spectrum together with its scalar index spectral in this scenario were found. 
By comparing the obtained results with the observations coming from the cosmic microwave background anisotropies, the constraints on the parameters space of the model and also its predictions are analyzed and discussed. 
\end{abstract}

\maketitle
\section{Introduction}\label{intro}
Inflation theory as an unavoidable part of modern cosmology attempts to describe the phenomena at the early time through the primordial perturbations coming from the quantum fluctuations. The scalar perturbations as the seed of the universe form the large-scale structures of the universe. Moreover, they are the main ones responsible for the temperature anisotropies of the cosmic microwave background (CMB) at the last scattering surface. On the other hand, monitoring the B-mode polarized CMB photons discloses the existence of primordial gravitational waves generated by the tensor perturbations of inflation \cite{Guth:1980zm,Linde:1981mu,Albrecht:1982wi,Lyth:1998xn}. In the standard cosmology, a single scalar field, so-that called inflaton, is the dominant matter of the universe during cosmic inflation so that it decays to the particles at the last step of inflation through the reheating process \cite{Kofman:1994rk,Shtanov:1994ce}. Non-minimal coupling (NMC) model is another broadly-used inflationary model considering a coupling between the Ricci scalar and the inflaton field coming from the quantum corrections of scalar field \cite{Lucchin:1985ip,Futamase:1987ua,Fakir:1990eg,Kallosh:2013maa,Edwards:2014poa,Shokri:2017mpl,Shokri:2019rfi,Shokri:2021rhy,Shokri:2021jxh,Upadhyay:2023wcq}. Besides the mentioned models, modified theories of gravity introduce different mechanisms in order to explain the early-time accelerating phase of the universe. For instance, in $f(R)$ gravity, modifying the Ricci scalar can describe the inflationary era instead of considering a scalar field \cite{Starobinsky:1980te,DeFelice:2010aj,Capozziello:2011et,Nojiri:2010wj,Ben-Dayan:2014isa,Sadeghi:2015nda,Motohashi:2012tt,Renzi:2019ewp}. Also, in the extended versions of teleparallel gravity, i.e. modified teleparallel gravity $f(T)$ and modified symmetric teleparallel gravity $f(Q)$, two other geometrical objects, tension and non-metricity, are responsible to demonstrate the inflationary epoch \cite{Bahamonde:2021gfp,BeltranJimenez:2019tme,Capozziello:2022tvv}. Apart from the inflationary criteria, these models could be also tested in the context of weak gravity conjecture (WGC), in particular, in the swampland region where the low-energy effective field theories are incompatible with string theory \cite{Vafa:2005ui,Obied:2018sgi,Agrawal:2018own,Blumenhagen:2018hsh,Shokri:2021iqp,Gashti:2022hey,Sadeghi:2023cxh}.

A wide range of inflationary literature has been dedicated to compere the predictions of single field models with the CMB observations. Consequently, some of them are excluded while some of them are still compatible with the Planck datasets \cite{Starobinsky:1980te,Barrow:1988xh,Martin:2016ckm}. As a shortcoming of these models, they don't predict any non-Gaussianity in their spectrum \cite{Chen:2010xka} while the future CMB observations propose the existence of the non-Gaussianity in the detected spectrum of the inflationary perturbations. In such a case, single field models will be put into question. To escape from this, constant-roll inflation 
has been suggested in which inflaton rolls down with a constant rate from the maximum point of the potential to the minimum point at the end of inflation as
\begin{equation}
\ddot{\varphi}=-(3+\alpha)H\dot{\varphi},
\label{1}
\end{equation}
where $\alpha$ is a non-zero parameter \cite{Motohashi:2014ppa,Motohashi:2017aob}. Deviation from the slow-roll approximation is also traced in the ultra slow-roll regime where we assume a non-negligible $\ddot{\varphi}=-3H\dot{\varphi}$. This class of inflationary models shows a finite value of the non-decaying mode in the spectrum of curvature perturbations \cite{Inoue:2001zt}. Although the solutions of ultra models are situated in the non-attractor phase of inflation, sometimes they present an attractor-like behaviour \cite{Pattison:2018bct}. Moreover, the large $\eta$ predicted by the ultra models can't guarantee to solve the $\eta$ problem introduced in supergravity \cite{Kinney:2005vj}. Assuming a fast rate of rolling at the beginning of inflation is comprehended as another kind of deviation from the slow-roll inflation \cite{Contaldi:2003zv,Lello:2013awa,Hazra:2014jka}. 

Recently, the idea of constant-roll has been engaged broadly for different inflationary models \cite{Odintsov:2017yud,Nojiri:2017qvx,Motohashi:2017vdc,Ito:2017bnn,Odintsov:2017hbk,Odintsov:2017qpp,Awad:2017ign,Lin:2019fcz,Motohashi:2019tyj,Odintsov:2019ahz,Shokri:2021aum,Shokri:2021iqp,Shokri:2021jxh,Shokri:2021rhy,Shokri:2021zqw, Herrera:2023ywx,Mohammadi:2022tmk,Ahmadi:2023qcw}. In this work, we focus on a type of inflationary models in which a complex scalar field is assumed as inflaton to drive inflation. Complex scalar fields in quantum and classical cosmology were first introduced in \cite{Khalatnikov:1992sj,Kamenshchik:1995ib} and then were developed in \cite{Amendola:1994xf,Scialom:1994uq,Kamenshchik:1997dmk}. Complex fields are comprehended as an alternative to the quintessence field in order to explain the late-time accelerating phase of the universe, so-that called dark energy (DE) \cite{Liu:2020bmp,Carvente:2020aae,Gu:2001tr}. Besides this, a complex scalar field can play the role inflaton in the early universe \cite{Yurov:2002nu}. Consequently, complex inflation naturally presents a graceful exit from the inflationary era with a very small number of \textit{e}-folds \cite{Yurov:2001ud}. Hence, a complex scalar field can be considered as an assistant field of a real inflaton in the context of hybrid inflation in order to terminate inflation \cite{Buchmuller:2014epa}. 

The main aim of the present manuscript is to study a constant-roll inflationary scenario by considering a complex scalar field as the scalar field driven inflation. To achieve this, we arrange the paper as follows. In Section \ref{review}, we review the foundation of complex scalar field theory and its dynamical equations, briefly. We investigate the complex inflation in the context of the constant-roll approach by using the accredited form of the angular speed $\dot{\theta}$ in Section \ref{cri}. In Section \ref{CP}, we present the cosmological perturbations of the complex constant-roll inflation. Also, we attempt to find the constraints on the parameter space of the model using the observational values of the spectral parameters. Conclusions and remarks are drawn in Section \ref{concl}. In the following, we chose units so that $c=\hbar= \kappa^2= 8\pi G=1.$

\section{Cosmology with a Complex Scalar Field}\label{review}

Let us start with the action of complex scalar theory \cite{Khalatnikov:1992sj,Kamenshchik:1995ib}
\begin{equation}
S=\int{d^{4}x\sqrt{-g}\bigg(\frac{R}{2}-\frac{1}{2}g^{\mu\nu}\partial_{\mu}\Psi^{*}\partial_{\nu}\Psi-V(|\Psi|)+\mathcal{L}_m\bigg)},
\label{2}   
\end{equation}
that considers a complex scalar field $\Psi$ minimally coupled to the Ricci scalar $R=g^{\mu\nu}R_{\mu\nu}$ in the presence of ordinary matter  described by a Lagrangian density $\mathcal{L}_m$. Here, we assume that the self-interacting potential $V$ just depends on the absolute value or amplitude of the complex scalar field \cite{Gu:2001tr} and the quantity $g$ corresponds to the determinant of the metric $g_{\mu\nu}$. 
We define the complex scalar field in terms of its amplitude $\varphi(x)$ and phase $\theta(x)$ by
\begin{equation}
\Psi(x)=\varphi(x)e^{i\theta(x)}.
\label{3}    
\end{equation}
This configuration allows us to describe both early and late-time accelerating phases of the universe only by using a complex field without any cosmological constant \cite{Liu:2020bmp,Gu:2001tr,Yurov:2002nu,Yurov:2001ud}. Moreover, the usefulness of using the $\varphi(x)$ and $\theta(x)$ will advantage the obtaining of  the motion equations (as well as the solutions) that relate the background variables, in particular the effective potential $V$ and the Hubble parameter $H$.
In this form, we can rewrite the action (\ref{2}) as
\begin{equation}
S=\int{d^{4}x\sqrt{-g}\bigg(\frac{R}{2}-\frac{1}{2}g^{\mu\nu}\partial_{\mu}\varphi\partial_{\nu}\varphi-\frac{1}{2}\varphi^{2}g^{\mu\nu}\partial_{\mu}\theta\partial_{\nu}\theta-V(\varphi)+\mathcal{L}_m\bigg)}.
\label{4} 
\end{equation}
By variation of the action with respect to the metric and then considering a spatially flat universe described by the Friedmann-Robertson-Walker (FRW) metric, the dynamical equations of the model are given as 
\begin{equation}
H^{2}\equiv\Big(\frac{\dot{a}}{a}\Big)^{2}=\frac{1}{3}\rho=\frac{1}{3}\Big(\rho_{m}+\frac{1}{2}\big(\dot{\varphi}^{2}+\varphi^{2}\dot{\theta}^{2}\big)+V(\varphi)\Big),
\label{5}    
\end{equation}
\begin{equation}
\frac{\ddot{a}}{a}=-\frac{1}{6}(\rho+3p)=-\frac{1}{3}\Big(\frac{1}{2}(\rho_{m}+3p_m)+\big(\dot{\varphi}^{2}+\varphi^{2}\dot{\theta}^{2}\big)-V(\varphi)\Big),
\label{6}    
\end{equation}
where the fluid filling the universe satisfies the continuity equation
\begin{equation}
\dot{\rho}+3H(\rho+p)=0,
\label{7}
\end{equation}
where $H$ is the Hubble parameter, $\rho=\rho_m+\rho_\Phi$ corresponds to the total energy density, $p=p_m+p_\Phi$ denotes the total pressure and the dot depicts the derivative with respect to cosmic time $t$. Here we have considered that   a spatially homogeneous complex scalar field $\Phi(t)$, such that $\Phi(t)=\varphi(t)e^{i\theta(t)}$ and the  tensor energy momentum $T^\text{o.m.}_{\mu\nu}$ associated to the ordinary matter Lagrangian can be described by a  perfect fluid of the form $T^\text{o.m}_{\mu\nu}=$diag$(-\rho_m,p_m,p_m,p_m)$ in which $\rho_m$ and $p_m$ denote the effective  energy density and pressure associated to ordinary matter, respectively.
Analogously,
we have assumed that the energy density and pressure of the complex scalar field are given by \cite{Gu:2001tr}
\begin{equation}
\rho_{\Phi}=\frac{1}{2}\big(\dot{\varphi}^{2}+\varphi^{2}\dot{\theta}^{2}\big)+V(\varphi),\hspace{1cm}p_{\Phi}=\frac{1}{2}\big(\dot{\varphi}^{2}+\varphi^{2}\dot{\theta}^{2}\big)-V(\varphi).
\label{8}    
\end{equation} 
By varying the action (\ref{4}) with respect to the scalar field $\varphi$, the equation of motion (EoM) for the scalar field $\varphi$ becomes
\begin{equation}
\ddot{\varphi}+3H\dot{\varphi}-\dot{\theta}^{2}\varphi+V'(\varphi)=0,
\label{9}    
\end{equation}
and for phase we have the motion equation and its solution (for angular speed $\dot{\theta}$) given by
\begin{equation}
\ddot{\theta}+(2\dot{\varphi}/\varphi+3H)\dot{\theta}=0,\,\,\,\,\,\,\Rightarrow\,\,\,\dot{\theta}\propto \frac{1}{\varphi^2\,a^3}.
\label{10}
\end{equation}
In the following we will consider that  the prime denotes the derivative with respect to the scalar field $\varphi$. 

In the context of  the scenario dominated  by the dark sector (DE) and in order to explain late time accelerating phase of the universe, the authors in Ref.\cite{Gu:2001tr} considered a non-relativistic matter with pressure $p_m=0$ together with a complex quintessence theory in which the angular velocity $\dot{\theta}$ is proportional to 
${a^{-3}\varphi^{-2}}$, see  Eq.(\ref{10}).
For a review of the complex field in the framework of DE, see Refs.\cite{Liu:2020bmp,Carvente:2020aae,Gu:2001tr}.

On the other hand, during the early universe the classical and quantum description of the a inflationary epoch from the complex scalar inflaton field  was studied in different articles \cite{Khalatnikov:1993vs,Amendola:1994xf,Kamenshchik:1997dmk,Lee:1994qb} and during the reheating and the primordial black hole production in \cite{Carrion:2021yeh}, see also Ref.\cite{Scialom:1996yd}. The importance of considering a complex scalar field  during this epoch lies in the fact that such fields, the non-Abelian multiplets of scalar fields and other fields 
arise naturally in the new theories of particle physics, string etc.
In particular in Ref.\cite{Yurov:2001ud} was utilized the Barrow's method to solve the inflationary dynamics associated to the complex field as a natural exist of inflation but with a small quantity of expansion under  an angular velocity given by the expression 
$\dot{\theta}=\frac{M}{\varphi^{2}},$
where the quantity $M$ corresponds to a constant. In particular, in the case  in which the constant $M=0$ the inflationary model reduces to,  inflation with a real scalar field. 
Despite the mentioned success, the complex inflaton field suffers from a problem that is the number of $e$-folds related to the expansion of the universe is very small. Hence, using a single complex field as inflaton field will be put into doubt \cite{Yurov:2001ud,Yurov:2002nu}. 

In the following we will assume that the ordinary matter Lagrangian $\mathcal{L}_m=0$ ($\rho=\rho_\Phi$ and $p=p_\Phi$), in order to study the constant roll inflation in the framework of the complex field. 
Thus, we find that 
the dynamical equations are given by 
\begin{equation}
3H^{2}=\frac{1}{2}\big(\dot{\varphi}^{2}+\varphi^{2}\dot{\theta}^{2}\big)+V(\varphi),\hspace{1cm}2\dot{H}=-\dot{\varphi}^{2}-\varphi^{2}\dot{\theta}^{2},
\label{11}    
\end{equation}
and
\begin{equation}
\ddot{\varphi}+3H\dot{\varphi}+V'(\varphi)=\varphi\dot{\theta}^2.
\label{12}    
\end{equation}
From Eq.(\ref{11}), we find that the effective potential can be written as 
\begin{equation}
V(\varphi)=3H^2+\dot{H}.
\label{13}
\end{equation}
In this form, knowing the Hubble parameter in terms of the scalar field together with the speed of the scalar field $\dot{\varphi}$, we can reconstruct the effective potential $V(\varphi)$ using Eq.(\ref{13}).

\section{Complex Constant-Roll Inflation}\label{cri}

To study the complex constant-roll inflation, we start with the constant-roll condition (\ref{1}) for the absolute value of the complex scalar field as
\begin{equation}
\ddot{|\Psi|}=-(3+\alpha)H\dot{|\Psi|},
\label{14}
\end{equation}
where as before the parameter $\alpha\neq0$. 
By using the definition of the complex scalar field (\ref{3}), the above constant-roll condition is reduced to the familiar form (\ref{1}).
From Eq.(\ref{11}) and considering  the definition $\dot{H}=\dot{\varphi}H'$, we find that the speed of the scalar field $\dot{\varphi}$ becomes
\begin{equation}
\dot{\varphi}=-H'\pm\sqrt{H'^2-\varphi^{2}\dot{\theta}^{2}}.
\label{15}  
\end{equation}
In order to find Eq.(\ref{15}), we have considered that the Hubble parameter is only function the scalar field $\varphi$  and not of  $\theta$ (or $\dot{\theta}$), since the solution given by Eq.(\ref{10}) allows eliminating the dependency of the Hubble parameter of this variable (see Eq.(\ref{11})). Thus, from Eq.(\ref{10}), we can write   
$\dot{H}=\varphi H'$ and then the relation given by Eq.(\ref{15}) becomes useful to determine $\dot{\varphi}$.

As we see, in order to find a real solution for $\dot{\varphi}$, we have that during the constant roll scenario it is necessary that $H'^2>\varphi^{2}\dot{\theta}^{2}$. Also, from Eq.(\ref{13}), the potential and the speed of the scalar field can be driven as
\begin{equation}
V(\varphi)=3H^{2}-H'^2\pm H'\sqrt{H'^2-\varphi^{2}\dot{\theta}^{2}}.
\label{16}    
\end{equation}
Now, taking the derivative of Eq.(\ref{15})  with respect to cosmic time $t$ and  using the constant-roll condition given by (\ref{1}), we obtain a differential equation for the  Hubble parameter in terms of the scalar field given by
\begin{equation}
-(3+\alpha)H=-H''\pm\left[\frac{H'H''-(\varphi^{2}\dot{\theta}^{2}/2)'}{\sqrt{H'^{2}-\varphi^{2}\dot{\theta}^{2}}}\right].
\label{17} 
\end{equation}
From Eq.(\ref{10}), the phase speed $\dot{\theta}$ can be written as $\dot{\theta}=\sqrt{w}/\varphi^2a^3$, with $w=constant>0$. Then, the speed of the real scalar field (\ref{15}) and the corresponding potential (\ref{16}) are rewritten as
\begin{equation}
\dot{\varphi}=-H'\pm\sqrt{H'^{2}-w/\varphi^2a^6},\hspace{1cm}V(\varphi)=3H^{2}-H'^2\pm H'\sqrt{H'^2-w/\varphi^2a^6}. 
\label{18}
\end{equation}
Moreover, we find that the differential equation for the Hubble parameter $H$ (\ref{17}) becomes
\begin{equation}
-(3+\alpha)H=-H''\pm\left[\frac{H'H''-(w/2\varphi^2a^6)'}{\sqrt{H'^{2}-w/\varphi^2 a^6}}\right]. 
\label{19}
\end{equation}
To obtain the analytical solutions for the background variables, we can consider that during inflation 
$1\gg w/\varphi^2 a^6H'^2$ with which 
\begin{equation}
-(3+\alpha)HH'\simeq-H'H''\pm[H'H''-(w/2\varphi^2a^6)'][1+\frac{w}{2\varphi^2a^6H'^2}+....],
\label{20}
\end{equation}
and for the speed of scalar field in this approximation  results
\begin{equation}
\dot{\varphi}\simeq-H'\pm H'\left[1-\frac{w}{2\varphi^2 a^6 H'^2}+...\right].
\label{21}
\end{equation}
By considering the positive sign and keeping the first term into the expansion given by Eq.(\ref{20}), we have the reduced differential equation 
\begin{equation}
(3+\alpha)HH'\simeq (\frac{w}{2\varphi^2a^6})',
\label{22}
\end{equation}
with the solution
\begin{equation}
H^2+C\simeq\frac{w_1}{\varphi^2 a^6},
\label{23}
\end{equation}
where we have defined  $w_1=w/(3+\alpha)$ and $C$ corresponds to an integration constant.  Also, in this case  the  quantity $\dot{\varphi}$ (\ref{21}) yields
\begin{equation}
\dot{\varphi}\simeq-\frac{w}{2\varphi^2 a^6 H'}=-\frac{w}{2w_1}\,\frac{(H^2+C)}{H'}=-\frac{(3+\alpha)}{2}\,\frac{(H^2+C)}{H'}.
\label{24}
\end{equation}
In the particular case in which the  integration constant $C=0$  and 
in order to find the Hubble parameter in terms of the scalar field, we take the derivative of Eq.(\ref{23}) with respect to scalar field results
\begin{equation}
H'=-H\left[\frac{1}{\varphi}+3\frac{H}{\dot{\varphi}}\right],
\label{25}
\end{equation}
where we have considered  that $H=\tilde{H}\dot{\varphi}$ where $\tilde{H}=a'/a$. Then, using Eq.(\ref{24}), the differential equation (\ref{25}) reduces to
\begin{equation}
H'\simeq-H\left[\frac{1}{\varphi}-\frac{6}{ (3+\alpha)}\,\frac{H'}{H}\right],
\label{26}
\end{equation}
and its solution is given by
\begin{equation}
 H(\varphi)=C_1\,[(3-\alpha)\varphi]^{-\beta},\,\,\,\,\mbox{with}\,\,\,\,\,\,\beta=\left(\frac{3+\alpha}{\alpha-3}\right),
 \label{27}
 \end{equation} 
where $C_1$ corresponds to another integration constant. Note that Eqs.(\ref{22}) and (\ref{25}) are equivalent, since we have only utilized  the solution given by Eq.(\ref{23}) (with $C=0$).

For the case in which the integration constant $C\neq0$, we can introduce the variable change  $\mathcal{H}^2=H^2+C$ into Eq.(\ref{23}) and using Eq.(\ref{24}),  we find the same differential equation  (\ref{26}) for the variable $\mathcal{H}$. Thus, we obtain that the solution  for the Hubble parameter as a  function of the scalar field in the case in which the integration constant $C\neq0$ yields
\begin{equation}
H(\varphi)=\sqrt{C_1^2[(3-\alpha)\varphi]^{-2\beta}-C},\label{new2}
\end{equation}
where as before $C_1$ denotes an integration constant and $\beta$ is defined by Eq.(\ref{27}). Clearly, when  the integration constant $C=0$, the solution given by Eq.(\ref{new2}) is reduced to the Eq.(\ref{27}). 

For simplicity, in the following we will focus our analysis  on the stage  in which the integration constant $C$ is set to zero. In this form, plugging Eq.(\ref{27}) into Eq.(\ref{24}), the scalar field as a function of the cosmological time  becomes
\begin{equation}
\varphi(t)=\frac{1}{(3-\alpha)}\left[\frac{C_1(3+\alpha)}{2}\,t+C_2\right]^{1/\beta},\,\,\,\,\,\,\mbox{with}\,\,\,\,\,\,\,\alpha\neq 3.
\label{28}
\end{equation}
The range of the validity for the scalar field during inflation can be determined from the condition $1\gg w/(\varphi^2a^6H'^2)$ results
\begin{equation}
\varphi^{-2}\gg \frac{(3+\alpha)}{\beta^2},\,\,\,\,\,\mbox{with which}\,\,\,\,\,\,\,
0>\varphi>-\frac{\vert\beta\vert}{\sqrt{(3+\alpha)}},\,\,\,\,\mbox{or}\,\,\,\,\,\,0<\varphi<\frac{\vert\beta\vert}{\sqrt{(3+\alpha)}},\label{cond}
\end{equation}
where we have used Eqs.(\ref{23}) and (\ref{27}), respectively.

Besides, we find that the phase speed $\dot{\theta}$ in terms of the scalar field becomes
\begin{equation}
\dot{\theta}=\frac{\sqrt{w}}{\varphi^2\,a^3}=\dot{\theta}_0\,\varphi^{-(\beta+1)},\,\,\,\,\,\,\mbox{where}\,\,\,\,\,\dot{\theta}_0=C_1(3+\alpha)^{1/2}\,(3-\alpha)^{-\beta}.\label{T1}
\end{equation}
Thus, combining Eqs.(\ref{28}) and (\ref{T1}) we obtain that the phase as a function of the time is given by 
\begin{equation}
\theta(t)=\theta_1\,\left[\frac{C_1(3+\alpha)}{2}\,t+C_2\right]^{-1/\beta}+C_3,
\label{T3}
\end{equation}
where the quantity $\theta_1$ is a constant and it  is defined as 
\begin{equation}
\theta_1=\frac{-2\beta\,\dot{\theta}_0}{C_1\,(3+\alpha)\,(3-\alpha)^{-(\beta+1)}},
\end{equation}
and $C_3$ corresponds to another integration constant.
Also, we find that the effective potential (\ref{13}) as a function of the scalar field  becomes
\begin{equation}
V(\varphi)=\left(\frac{3-\alpha}{2}\right)C_1^2[(3-\alpha)]^{-2\beta}\,\varphi^{-2\beta}=V_0\,\varphi^{-2\beta},
\label{29}
\end{equation}
where we have defined $V_0=\frac{(3-\alpha)}{2}C_1^2[(3-\alpha)]^{-2\beta}$. 
Here we observe that from the approximation $w(\varphi\,a^3\,H')^{-2}\ll 1$, as much as the solution for the Hubble parameter $H(\varphi)$ and  the reconstructed potential $V(\varphi)$ do not depend on the parameter $w$ associated to the angular speed $\dot{\theta}$. 
This is due to the fact that we have considered the first term into the expansion (\ref{20}) and then Eq.(\ref{24}) does not depend on $w$. In the situation in which we keep the first and second term of the expansion given by Eq.(\ref{20}), the differential equation (\ref{22}) is modified to
\begin{equation}
\left[\frac{2(3+\alpha)}{w}\right]\,HH'\simeq \left(\frac{1}{\varphi^2a^6}\right)'\left[1+\frac{w}{2\varphi^2a^6H'^2}\right]-\frac{H''}{\varphi^2a^6H'},
\label{22b}
\end{equation}
and together with the equation $\dot{\varphi}\simeq -w/(2\varphi^2 a^6 H')$ obtained of (\ref{21}) are part of the system  to be solved (recall that $H=(a'/a)\dot{\varphi}$). Thus, in order to solve (numerically) this coupled system of differential equations we would need to specify the value of the parameter $w$.  

\begin{figure*}[!hbtp]
     \centering	\includegraphics[width=0.6
\textwidth,keepaspectratio]{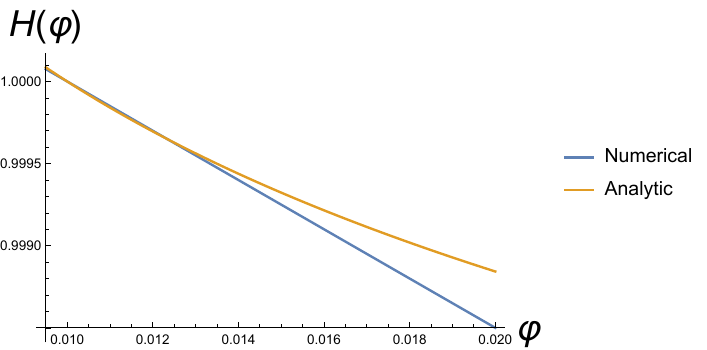}
\caption{The plot shows  the comparison between the analytic expression given by Eq.(\ref{27}) and the numerical solution  for the Hubble parameter $H(\varphi)$ in terms of the scalar field $\varphi$. Here we have used the positive signs of Eqs.(\ref{18}) and (\ref{19}) together with the values of  $\alpha=-2.99$, $w=0.01$ and $C_1=0.99$,  respectively.}
\label{fig1A}
\end{figure*}
The Fig.\ref{fig1A} shows the comparison between the analytic solution given by Eq.(\ref{27}) where the approximation $1\gg w/\varphi^2a^6H'^2$ is used and the numerical solution for the Hubble parameter as a function of the scalar field. Here we have utilized the values $\alpha=-2.99$, $w=0.01$ and $C_1=0.99$. We note that the comparison is suitable between the numerical solution and the analytical solution inside of the   validity range   defined by  Eq.(\ref{cond}) for the scalar  field. 

Besides, the number of $e-$folds in this case results
\begin{equation}
N=-\int_{t_*}^{t_f}H dt=-\int_{\varphi_*}^{\varphi_f}
\frac{H}{\dot{\varphi}}d\varphi=\frac{2w_1}{w}\int_{\varphi_*}^{\varphi_f}\frac{H'}{H}d\varphi=-\frac{2}{(\alpha-3)}\,\ln\left[\frac{\varphi_f}{\varphi_*}\right].
\label{30}
\end{equation}

In the following, the subscripts $f$ and $*$ are used to describe to the epoch the end of inflation and when the cosmological scales exit the horizon, respectively.

By assuming that the end of inflation occurs when the slow roll parameter $\epsilon_1\simeq\frac{1}{2}(V'/V)^2=1$, then we find that the value of the scalar field at the end of inflation is given by 
\begin{equation}
\epsilon_1(\varphi=\varphi_f)\simeq\frac{2\beta^2}{\varphi_f^2}=1,\,\,\,\,\,\,\Rightarrow\,\,\,\,\,\,\,\varphi_f=\pm \sqrt{2}\beta.
\label{31}
\end{equation} 

In this way, we obtain that the value of the scalar field when the cosmological scales exit the horizon yields
\begin{equation}
\varphi_*=-\sqrt{2}\,\beta\,e^{(\alpha-3)N/2},
\label{32}
\end{equation}
where we have considered  the negative sign of $\varphi_f$ (see Eq.(\ref{31})), since  as we will see later the parameter $\beta$ is a negative quantity.

\section{Cosmological Perturbations }\label{CP}
In this section, we will analyze the scalar and tensor perturbations, in which  the former will be characterized by considering the longitudinal gauge on the metric.  Thus,  by considering the longitudinal gauge in the perturbed FRW  
 metric we write
\begin{equation}
ds^2=(1+2\Phi)dt^2-a^2(t)(1-2\Phi)\delta_{ij}dx^idx^j,
\label{n1}
\end{equation}
where the function $\Phi=\Phi(t,\bf{x})$ corresponds to the Bardeen's gauge invariant variable \cite{Bardeen:1980kt}. Assuming the spatial dependence  $e^{i\bf{kx}}$, where the quantity $k$ is the wavenumber, each Fourier mode satisfies the following perturbed equations of motion given by 
\begin{equation}
\dot{\Phi}+H\Phi=\frac{1}{2}\left[\dot{\varphi}\delta\varphi+\varphi^2\dot{\theta}
\delta\theta\right],
\label{n2}
\end{equation}
\begin{equation}
\delta\ddot{\varphi}+3H\delta\dot{\varphi}+\left[\frac{k^2}{a^2}+V''-\dot{\theta}^2\right]\delta\varphi-2\varphi\dot{\theta}\delta\dot{\theta}=4\dot{\varphi}\dot{\Phi}-2V'\Phi,
\label{n3}
\end{equation}
and
\begin{equation}
\delta\ddot{\theta}+\left[3H+\frac{2}{\varphi}\dot{\varphi}\right]\delta\dot{\theta}+ \frac{k^2}{a^2}\,\delta\theta-\frac{2\dot{\varphi}}{\varphi}\delta\dot{\varphi}=4\dot{\theta}\dot{\Phi}.
\label{n4}
\end{equation}
Here, the quantities $\delta\varphi$ and $\delta \theta$ correspond to the gauge invariant fluctuation 
variables associated to the respective fields $\varphi$ and $\theta$, respectively. In order to establish that the quantities $\delta\phi$ and $\delta\theta$ are gauge invariant fluctuation variables,  we can consider an  infinitesimal gauge transformation on the coordinates $\tilde{x^{\nu}}=x^{\nu}+\delta x^\nu$ together with the most generic perturbed metric; $ds^2=a^2[(1+2\Phi)d\tau^2-2\partial_iBd\tau dx^{i}-[(1-2\psi)\delta_{ij}+(\partial_i\partial_j-(1/3)\delta_{ij}\nabla^2)E]dx^{i}dx^{j}]$, with $\tau$ the conformal time and  a detailed explication see Refs. \cite{Mukhanov:1990me,Riotto:2002yw}. Since $\varphi$ and $\theta$ are scalar fields (amplitude and phase), these satisfy the transformation laws  for gauge invariant (GI);   $\delta\varphi_{GI}=\delta\varphi-\varphi'(E'/2-B)$ and $\delta\theta_{GI}=\delta\theta-\theta'(E'/2-B)$, where a prime now indicates differentiation wrt to the conformal time \cite{Mukhanov:1990me,Riotto:2002yw}. Thus, from the longitudinal gauge on the metric given by Eq.(\ref{n1}), we have $B=E=0$ and $\Phi=\psi$ and then the quantities $\delta\varphi$ and $\delta\theta$ are gauge invariant fluctuation variables. 

By considering large scale perturbations in which $k\ll aH$ and neglecting the term $\dot{\Phi}$  and those terms which include second  order time derivative, then the above equations of motion  reduce to
\begin{equation}
H\Phi\simeq\frac{1}{2}\left[\dot{\varphi}\delta\varphi+\varphi^2\dot{\theta}\delta\theta\right],
\label{n5}
\end{equation}
\begin{equation}
3H\delta\dot{\varphi}+V''\delta\varphi\simeq-2V'\Phi,
\label{n6}
\end{equation}
and
\begin{equation}
3H\delta\dot{\theta}\simeq 0,\,\,\,\,\,\Rightarrow\,\,\,\delta\theta =\mathcal{C}.
\label{n7}
\end{equation}
Here we have considered that as we need the non-decreasing modes on large scale in our model, which are determined to be weakly time-dependent \cite{Polarski:1992dq,Polarski:1994rz}, then  we can consistently ignore the term associated to $\dot{\Phi}$, including those involving two time derivatives. In relation to the existence of growing and decaying modes is attributed  to the the observation outlined  in Ref.\cite{Starobinsky:1982ee},  with  a more detailed explication provided in  Ref.\cite{Polarski:1992dq}.

Thus, we find that the Eq.(\ref{n6}) can be written as
\begin{equation}
\delta\varphi'+\frac{1}{3H\dot{\varphi}}\left[V''+\frac{V'}{H}\dot{\varphi}\right]\delta\varphi+\left(\frac{V'\varphi^2\dot{\theta}}{3H\dot{\varphi}}\right)\mathcal{C}=0,
\label{n8}
\end{equation}
where we have used $\delta\dot{\varphi}=\dot{\varphi}\delta\varphi'$. 

\begin{figure*}[!hbtp]
     \centering	\includegraphics[width=0.6
\textwidth,keepaspectratio]{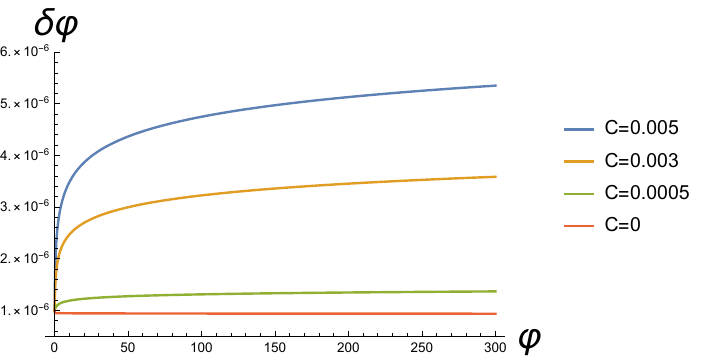}
\caption{The plot shows  the numerical solution for the gauge invariant fluctuation variable $\delta\varphi$ versus the scalar field $\varphi$ given by Eq.(\ref{n8}), for different values of the constant $\mathcal{C}$. In this plot we have utilized the values $C_1=0$, $C_2=0$ and $\alpha=-2.99$. }
\label{fig2A}
\end{figure*}

We note that the relation between the real variables  $\delta\varphi$ and $\delta\theta$ with the variation of the complex field $\delta\Psi$ from Eq.(\ref{3}) becomes $\delta\Psi=(\Psi/\varphi)[\delta\varphi+ i\,\delta\theta]$. Under the large scale  approximation,  we find that the relation between the Bardeen's variable  $\Phi$ and the variation of the complex field $\delta\Psi$ can be written as
$2H\Phi\simeq\delta\Psi[(\varphi/\Psi)\dot{\Psi}-i\,\dot{\theta}](\varphi/\Psi)-\mathcal{C}[\dot{\theta}(1-\varphi^2)+i\,(\varphi/\Psi)\dot{\Psi}]$, where we have utilized Eqs.(\ref{n5}) and (\ref{n7}), respectively. Thus, this relation indicates that the metric perturbation is determined when the evolution of the  fields are known.

In order to find an analytical solution to the 
gauge invariant variable $\delta\varphi$, we consider for the integration constant as $\mathcal{C}=0$. In this way,  we find that the solution of Eq.(\ref{n8}) can be written as
\begin{equation}
\delta\varphi=\mathcal{C}_0\,\exp\left[-\int_\varphi^{\varphi_f}\, \frac{1}{3H\dot{\varphi}}\left(V''+\frac{V'}{H}\dot{\varphi}\right)d\varphi\right],\label{n9}
\end{equation}
where $\mathcal{C}_0$ corresponds to an integration constant. Therefore, the variable $\Phi$ becomes 
\begin{equation}
\Phi\simeq\frac{1}{2}\frac{\dot{\varphi}}{H}\,\mathcal{C}_0\,\exp\left[-\int_\varphi^{\varphi_f}\, \frac{1}{3H\dot{\varphi}}\left(V''+\frac{V'}{H}\dot{\varphi}\right)d\varphi\right].
\label{n10}
\end{equation}
Thus, from Eq.(\ref{24}), we find that Eq.(\ref{n9}) can be rewritten as
\begin{equation}
\delta\varphi\simeq\,\mathcal{C}_0\exp\left[\int_\varphi^{\varphi_f}\, \frac{2w_1H'}{3wH(H^2+C)}\left(V''-\frac{w(H^2+C)V'}{2w_1H\,H'}\right)d\varphi\right],\label{n11}
\end{equation}
and analogously for $\Phi$, we get
\begin{equation}
\Phi\simeq-\mathcal{C}_0\,\left(\frac{w(H^2+C)}{4w_1H\,H'}\right)\,\exp\left[\int_\varphi^{\varphi_f}\, \frac{2w_1H'}{3wH(H^2+C)}\left(V''-\frac{w(H^2+C)V'}{2w_1H\,H'}\right)d\varphi\right].
\label{n12}
\end{equation}
The Fig.\ref{fig2A} shows the numerical solution given by Eq.(\ref{n8}) for the fluctuation variable $\delta\varphi$ in terms of the scalar field $\varphi$, for different values of the constant $\mathcal{C}$ defined by Eq.(\ref{n7}). Here we have used the values $C_1=1$, $C_2=0$ and $\alpha=-2.99$. In order to write down values for the numerical solution of Eq.(\ref{n8}), we have utilized Eqs.(\ref{27}),  (\ref{28}), (\ref{T1})  and (\ref{29}) for the background variables. From this plot, we observe  that for values of integration constant $\mathcal{C}\ll 1$, the numerical solutions of $\delta\varphi (\varphi)$ exhibit behaviors closely resembling those observed when $\mathcal{C}=0$.  Hence, given that the gauge-invariant fluctuation variable $\delta\varphi$ remains significantly smaller than unity ($\delta\varphi\ll 1$), it is pertinent to consider the scenario where $\mathcal{C}=0$ as a suitable approximation. 

Following Refs.\cite{Starobinsky:1994mh,Langlois:2008mn,Starobinsky:2001xq}, we assume that gauge invariant density fluctuation $\delta\rho^c/\rho=\delta_H$ is defined as $\delta_H=-(2/3)(k/aH)^2\Phi$  and considering that the curvature perturbation due to  primordially adiabatic fluctuation  $\Phi=[1+2/(3(1+p/\rho))]^{-1}\mathcal{C}_0$, then from Eq.(\ref{n11}) we find that the density perturbation at the horizon crossing in which $k=aH$, results
\begin{equation}
\delta_H=f\left(\delta\varphi\,\,\,\exp\left[-\int_\varphi^{\varphi_f}\, \frac{2w_1H'}{3wH(H^2+C)}\left(V''-\frac{w(H^2+C)V'}{2w_1H\,H'}\right)d\varphi\right]\right) \Big\vert_{\varphi=\varphi_*}=f\left(\delta\varphi\,e^{{-\mathcal{F}}(\varphi)}\right)\Big\vert_{\varphi=\varphi_*},
\label{n13}
\end{equation}
where the function $\mathcal{F}(\varphi)$ is defined as
\begin{equation}
\mathcal{F}(\varphi)=\int_\varphi^{\varphi_f}\, \frac{2w_1H'}{3wH(H^2+C)}\left(V''-\frac{w(H^2+C)V'}{2w_1H\,H'}\right)d\varphi.
\label{n14}
\end{equation}
Further, the quantity $f$ is a constant and
in the specific case associated to the  radiation domination $f$ takes value $f=4/9$ and for matter domination we have $f=2/5$. Here the constant $f=(2/3)[1+2/(3(1+p/\rho))]^{-1}$ is introduced by normalization at the second crossing after inflation during the radiation or matter domination Refs.\cite{Starobinsky:1994mh,Langlois:2008mn,Starobinsky:2001xq}. Also the minus sign can be omitted, since this sign may be absorbed  into the stochastic variable $\delta\varphi$. Thus, the large-angular-scale anisotropy of background due to the Sachs-Wolfe effect is well defined as $\delta T/T=\Phi/3$ \cite{Sachs:1967er}. 

Besides, the fluctuations $\delta\varphi$ are generated by small scale perturbations and then they can be assumed as free massless scalar field
in which are described by independent random variables such that at the horizon crossing we have $\delta\varphi= H/2\pi$, see e.g. Refs.\cite{Lyth:1998xn,Liddle:2000cg}. By considering that the scalar power spectrum
$\mathcal{P_S}=(25/4)\delta_H^2$ \cite{Bassett:2005xm}, then we find that the spectrum at the horizon crossing can be written as
\begin{equation}
\mathcal{P_S}(\varphi=\varphi_*)=\left(\frac{25}{4}\right)\,\left(\frac{f}{2\pi}\right)^2\,\left[H^2\,e^{-2\mathcal{F}(\varphi)}\right]\Big\vert_{\varphi=\varphi_*}.
\label{n15}
\end{equation}
Further the scalar spectral index $n_s$ is defined as
\begin{equation}
n_s-1=\frac{d\ln \mathcal{P_S} }{d\ln k},
\label{n16}
\end{equation}
in which the wavenumber $k$ is related to the number of $e$-folds $N$ by the relationship  $d\ln k\simeq d\,N$. Thus,  from Eq.(\ref{n15}) we find that the index $n_s$ can be written as
\begin{equation}
n_s\simeq 1+2\frac{\dot{\varphi}}{H}\,\left[\frac{H'}{H}-\mathcal{F}\,'(\varphi)\right]
\simeq 1-\left(\frac{w}{w_1}\right)\,\frac{(H^2+C)}{H\,H'}\,\left[\frac{H'}{H}-\mathcal{F}\,'(\varphi)\right].
\label{n17}
\end{equation}

Besides,  it is well known that the production of tensor perturbations during the inflationary epoch would produce gravitational wave. In this context, the spectrum of the tensor perturbations $\mathcal{P_T}$ after Hubble exist can be approximated to \cite{Bassett:2005xm,Liddle:2000cg}
\begin{equation}
\mathcal{P_T}\simeq\, 4\left(\frac{H}{\pi}\right)^2\Big\vert_{\varphi=\varphi_*}.
\end{equation}

Also, an important observational parameter corresponds to the tensor to scalar ratio which is defined as
\begin{equation}
r=\frac{\mathcal{P_T}}{\mathcal{P_S}},
\end{equation}
and in our case this ratio can be written as
\begin{equation}
r\simeq\frac{64}{25}\frac{1}{f^2}e^{2\mathcal{F}(\varphi)}.
\end{equation}

For our model, from Eq.(\ref{23}), we find that the function 
$\mathcal{F}(\varphi)$ defined by Eq.(\ref{n14}) yields
\begin{equation}
\mathcal{F}(\varphi)=\frac{\beta(\alpha-3)}{3}\Bigg\{\ln(\varphi)+\frac{\beta(1+2\beta)}{\varphi^2}\frac{w_1}{w}\bigg(1+{}_2F_1\Big[1,-\frac{1}{\beta},\frac{-1+\beta}{\beta},-\frac{C(\varphi(\alpha-3))^{2\beta}}{C_1^2}\Big]\bigg)\Bigg\}\Bigg|^{\varphi}_{\varphi_{f}},
\nonumber
\end{equation}
where the quantity ${}_2F_1$ corresponds to the hypergeometric function. In the particular case in which the integration constant  $C=0$, the above function reduces to
\begin{equation}
\mathcal{F}(\varphi)=\frac{\beta(3-\alpha)}{3}\Bigg\{\ln(\varphi)+\frac{\beta(1+2\beta)}{\varphi^2}\frac{w_1}{w}\Bigg\}\Bigg|^{\varphi}_{\varphi_{f}}.\label{FF}
\end{equation}

By considering Eqs.(\ref{n17}) and (\ref{FF}) we obtain the that scalar spectral index $n_s$ can be written as 
\begin{equation}
n_s(\varphi=\varphi_*)=n_{s_*}\simeq -(2+\alpha)+\frac{(\alpha^2-9)}{3}\left[1-\frac{2\beta(1+2\beta)}{\varphi^2}\frac{w_1}{w}\right]\Bigg|_{\varphi_{*}}.
\label{nsa}
\end{equation}
Also, the tensor-to-scalar ratio of the model is given by
\begin{equation}
r_*\simeq\frac{64}{25}\frac{1}{f^2}\exp\bigg[\frac{2\beta(3-\alpha)}{3}\Big(\ln(\frac{\varphi}{2\sqrt{\beta}})+\frac{1+2\beta}{3+\alpha}\big(\frac{\beta}{\varphi^2}-\frac{1}{4}\big)\Big)\bigg]\Bigg|_{\varphi_{*}}.
\label{FFF}
\end{equation}

Thus, considering that the scalar spectral index at the horizon crossing takes the value $n_s{_*}=0.967$ and using Eqs.(\ref{31}) and (\ref{32}), we find numerically  that  the parameter $\alpha$ has a negative value given  by
\begin{equation}
\alpha\simeq\,-2.989.\label{alp}
\end{equation}
  Here we have considered that at the crossing the number of $e-$ folds $N=60$. 

Besides, from the scalar power spectrum defined by Eq.(\ref{n15}) and considering that at the crossing this quantity is $\mathcal{P_S}(\varphi=\varphi_*)\simeq 2.2\times 10^{-9}$ together with $\alpha=-2.989$,  we find that the integration constant $C_1=\pm \,1.02\times 10^{-4}$. However, as the Hubble parameter is a positive quantity we only have to consider $C_1= \,1.04\times 10^{-4}$. This value of $C_1$ suggests that the quantity $V_0=\frac{(3-\alpha)}{2}C_1^2(3-\alpha)^{-2\beta}$  (amplitude of the potential)  associated to the reconstructed potential $V(\varphi)=V_0\varphi^{-2\beta}$ becomes $V_0\simeq 3.1\times 10^{-8}$, when $\alpha=-2.989$ and $C_1\simeq 10^{-4}$. In this form, we find that   the value of the effective potential at the end of inflation results $V(\varphi=\varphi_f)\simeq 3.03\times10^{-8}$ (in units of $M_p^4$, where $M_p$ the Planck mass ) and at the crossing in which the number of $e-$folds $N=60$ the potential $V(\varphi=\varphi_*)\simeq 5.88\times 10^{-8}$ (in units of $M_p^4$). In this context, we obtain that the energy density of the universe during inflation in our model becomes $\rho\sim V\sim \mathcal{O}(10^{-7})\sim\mathcal{O}(10^{-8})$ $M_p^4$ and this energy scale is similar to that obtained  in  different inflationary models during the early universe, see e.g.,  \cite{Guth:1980zm,Linde:1981mu,Albrecht:1982wi,Bassett:2005xm}.
As before we have used that the number of $e-$folds $N=60$ and also we have considered the value $f=4/9$.

\begin{figure*}[!hbtp]
     \centering	\includegraphics[width=0.435\textwidth,keepaspectratio]{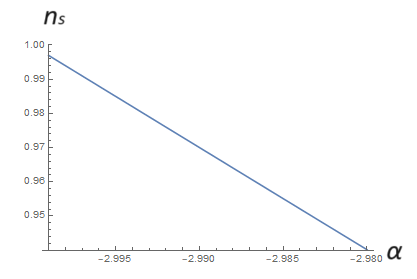}
\includegraphics[width=0.43\textwidth,keepaspectratio]{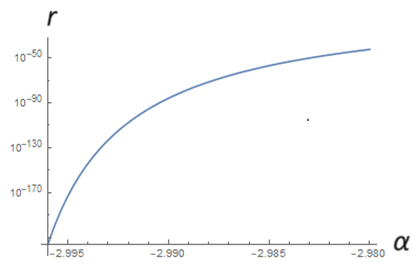}\vspace{0.5cm}
\caption{Observational parameters: The left panel shows the behavior of the scalar spectral index versus the parameter $\alpha$ and the right panel shows the tensor to scalar ratio $r$ (logarithmic scale) versus the parameter $\alpha$.  From the left panel we observe that the value $n_s=0.967$ for the scalar spectral index corresponds to $\alpha=-2.989$. Additionally,  from the right panel we note that the tensor to scalar ratio $r\sim 0$ for values of $\alpha$ in which $n_s\simeq 0.97$. Here we have used that at the crossing the number of $e-$folds $N=60$ and $f=4/9$. }
\label{fig1a}
\end{figure*}

Also, from Eq.(\ref{FFF}) we find that the tensor to scalar ratio $r_*\sim10^{-78}$, when we consider the values of $N=60$, $f=4/9$,  together with  the value given by Eq.(\ref{alp}) for the parameter $\alpha$. In this sense, we find that our model predicts that the tensor to scalar ratio ratio $r\simeq 0$, see right panel of Fig.\ref{fig1a}.

In Fig.\ref{fig1a} we show the evolution of the  observational parameters $n_s$ and $r$ versus the parameter $\alpha$  associated to Eq.(\ref{14}). The left panel shows that the observational value  for the scalar spectral index $n_s=0.967$ takes place for the value $\alpha\simeq-2.989$. In order to write down the scalar spectral index $n_s$ as a function of the parameter $\alpha$, we consider Eq.(\ref{nsa}) together with the Eq.(\ref{32}) in which the number of $e-$folds $N=60$. The right panel shows the evolution of the tensor to scalar ratio $r$ (logarithmic scale) in terms of the parameter $\alpha$. Also, in order to write down the observational parameter $r$ as a function of $\alpha$, we consider Eqs.(\ref{31}), (\ref{32}) and (\ref{FFF}), where we have assumed $N=60$ and the parameter $f=4/9$, respectively. From this panel, we note that the tensor to scalar ratio $r\simeq0$, when the observational parameter  $n_s\sim 0.97$.
On the other hand, in order to give an appropriate comparison and distinction  between the real inflaton field and the complex field, we have to consider the situation in which the parameter $\alpha>-3$. In this context, when we analyze  the case in which $\alpha>-3$ for a real field, see Ref.\cite{Motohashi:2014ppa}, the reconstructed potentials in terms of the scalar field  are an exponential potential associated to power-law inflation \cite{Abbott:1984fp} and a hyperbolic cosine  potential which is similar to that found in Ref.\cite{Barrow:1994nt} (inflationary solutions).  For the complex field we have found that the reconstructed potential corresponds to a power-law potential $V(\varphi)\propto \varphi^{-2\beta}$ becoming different to the real field. Here the evolution of the scalar field on the reconstructed potential in the framework of complex field becomes similar to the hyperbolic cosine  potential, since  both potentials present a minimum at $\varphi=0$, and the evolution of the field takes place from large-$\varphi$ towards the minimum. In relation to the observational parameters and in particular from  the scalar spectral index $n_s$, it is found that the parameter $\alpha$ takes the value $\alpha=-3.02$  when the index $n_s=0.96$ \cite{Motohashi:2014ppa}. In our case using a complex field we have obtained the value $\alpha=-2.989$ and it suggests that both constrains on the parameter $\alpha$ are similar from observational parameter $n_s$. However, the analysis realized assuming a real inflaton field predicts $\alpha<-3$ and from a complex inflaton field suggests $\alpha>-3$ from the scalar spectral index. 

\section{Conclusions and Remarks}\label{concl}

In the present manuscript, we have studied the complex inflation in which a complex field as inflaton is the main responsible to drive inflation. We have worked with an expression for the angular velocity $\dot{\theta}\propto\varphi^{-2}\,a^{-3}$ determined  from the motion equation for the phase $\theta$  that leads to some interesting inflationary solution.

By applying the constant-roll condition on the absolute value of the complex scalar field, we have found a specific form for the Hubble parameter $H$ and of the effective potential as a function of the scalar field,  by assuming that during inflationary era the term $w(\varphi\,a^3\,H')^{-2}\ll 1$. Here the solutions for the Hubble parameter and the effective potential have a dependency power-law type with the scalar field $\varphi$. Also, we have found that the evolution of the phase as a function of the cosmological time scale as $\theta(t)\propto t^{-1/\beta}$. In this context, we have noted that in the approximation $w(\varphi\,a^3\,H')^{-2}\ll 1$ the parameter related to the angular speed $w$ is not present in the solutions of $H(\varphi)$ and $V(\varphi)$, then this parameter can not be constrained.
Besides, from the condition $w(\varphi\,a^3\,H')^{-2}\ll 1$ we have found different ranges of validity for the scalar field, in which an upper bound is obtained for positive values of $\varphi$ or a lower bound for negative values of $\varphi$, see Eq.(\ref{cond}).  In this form, we have utilized   these ranges for the scalar field to constraint
the different results found on  parameter-space from the observational parameters.

Additionally, we investigated the corresponding scalar perturbations where we have obtained the perturbed equations of motion using the longitudinal gauge.
By considering large scale perturbations and neglecting 
some terms we have reduced the equations system and then we have found solutions for the Bardeen's variable and the perturbation associated to the scalar field. A general relation   for the scalar power spectrum and the scalar spectral index $n_s$ are given by Eqs.(\ref{n15}) and (\ref{n17}), respectively. In particular from the background solutions for the Hubble parameter and the effective potential in terms of the scalar field we have  found analytical quantities for the  observational parameters such as; scalar spectral index and the tensor to scalar ratio.

 By comparing the obtained results with the datasets coming from CMB anisotropies, we have attained the observational constraints on the parameters space of the model, in particular, the constant-roll parameter $\alpha$ see Eq.(\ref{alp}) and the integration constant associated to the Hubble parameter $C_1\sim\mathcal{O}(10^{-4})$.

Also, from Eq.(\ref{FFF}) we have obtained  that the tensor to scalar ratio $r_*\sim \mathcal{O}(10^{-78})$, when we consider the values of $N=60$ and $\alpha=-2.989$. This value for the tensor to scalar ratio suggests that our analysis done in the framework of  constant roll   with a complex inflaton, predicts a ratio $r\sim 0$, see Fig.\ref{fig1a}.  
Additionally, from Fig.\ref{fig1a} (left panel) we have ratified that the observational parameter $n_s\sim \mathcal{O}(1)$ when the parameter $\alpha\sim -3$.

In relation to the condition give by Eq.(\ref{1}), we see the   difficult to understand how
the scalar field could begin its movement on the reconstructed potential given by  (\ref{29}) with the 
appropriate initial condition $\ddot{\varphi}=-(3+\alpha)H\varphi$, in which the potential satisfies the relation $V'(\varphi)=\alpha H\dot{\varphi}+\varphi\dot{\theta}^2$. 
In this sense, the study of this issue concerning the consequences of the constant roll initial condition from the dynamics of the inflaton 
and how would be
the special features that would be a consequence this initial conditions in our scenario would be a very interesting subject
for future research concerning our model.

\bibliographystyle{ieeetr}
\bibliography{biblo}
\end{document}